\documentclass{article}

\usepackage[english]{babel}
\usepackage[a4paper,top=2cm,bottom=2cm,left=3cm,right=3cm,marginparwidth=1.75cm]{geometry}

\usepackage{amsmath}
\usepackage{graphicx}
\usepackage[colorlinks=true, allcolors=blue]{hyperref}
\usepackage{authblk}
\usepackage{float}
\usepackage{multirow} 
\usepackage{inconsolata} 
\usepackage{listings} 
\usepackage{siunitx}
\usepackage{xcolor} 
\usepackage[backend=biber,style=numeric-comp,sorting=none,maxbibnames=5]{biblatex}
\usepackage{csquotes}

\DeclareFieldFormat[article]{volume}{\mkbibbold{#1}}
\DeclareFieldFormat[article]{year}{(#1)}
\renewbibmacro{in:}{}

\title{{\tt dynsight}: an Open Python Platform for Simulation and Experimental Trajectory Data Analysis}

\author[1]{Simone Martino}
\author[1*]{Matteo Becchi}
\author[1,2]{Andrew Tarzia}
\author[1,3]{Daniele Rapetti}
\author[1*]{Giovanni M. Pavan}

\affil[1]{Department of Applied Science and Technology, Politecnico di Torino, Corso Duca degli Abruzzi 24, 10129 Torino, Italy}
\affil[2]{Present address: School of Chemistry, University of Birmingham, Edgbaston, Birmingham, B15 2TT UK}
\affil[3]{Present address: Scuola Internazionale Superiore di Studi Avanzati (SISSA), Trieste, Italy}
\affil[*]{\small{Corresponding authors: giovanni.pavan@polito.it, matteo.becchi@polito.it}}
\date{}

\begin{document}
\maketitle

\begin{abstract}
The study of complex many-body systems {\it via} analysis of the trajectories of the units that dynamically move and interact within them is a non-trivial task. The workflow for extracting meaningful information from the raw trajectory data is often composed of a series of interconnected steps, such as, (i) identifying and tracking the constitutive objects/particles, resolving their trajectories (e.g., in experimental cases, where these are not automatically available as in typical molecular simulations), (ii) translating the trajectories into data that are easier to handle/analyze by using well suited descriptors, and (iii) extracting meaningful information from such data. Each of these different tasks often requires non-negligible programming skills, the use of various types of representations or methods, and the availability/development of an interface between them. Despite the considerable potential that new tools contributed to each of these individual steps, their integration under a common framework would decrease the barrier to usage (especially by diverse communities of users), avoid fragmentation, and ultimately facilitate the development of new approaches in data analysis. To this end, here we introduce {\tt dynsight}, an open Python platform that streamlines the extraction and analysis of time-series data from simulation- or experimentally-resolved trajectories. {\tt dynsight} simplifies workflows, enhances accessibility, and facilitates time-series and trajectories data analysis offering a useful tool to unraveling the dynamic complexity of a variety of systems (or signals) across different scales. {\tt dynsight} is open source (\href{https://github.com/GMPavanLab/dynsight}{github.com/GMPavanLab/dynsight}) and can be easily installed using {\tt pip}.
\end{abstract}

\section{Introduction}
The analysis of complex systems is often non-trivial. Typical examples include systems composed of many units (atoms, molecules, objects, particles, etc.) that interact dynamically with each other and evolve collectively over time towards emergent functions.
These systems typically exhibit rich and complex behaviors that can be examined at different levels of detail and/or across various length and time scales to be understood.
However, collecting all the needed data and analyzing them across multiple scales often requires considerable expertise, proficiency in the use of different methods and, eventually, programming activities that may hinder understanding in diverse settings.

In the past decades, trajectory analysis in molecular simulations has greatly benefited from the development of software, such as {\tt MDAnalysis}\supercite{michaud2011mdanalysis, gowers2019mdanalysis}, {\tt Ambertools}\supercite{Case2023ambertools}, {\tt PLUMED}\supercite{Bonomi2009Plumed, Tribello2014Plumed2}, {\tt VMD}\supercite{Humphrey1996VMD}, {\tt OVITO}\supercite{Stukowski2010Ovito}, {\tt Chemiscope}\supercite{fraux2020chemiscope} (to name a few). These software have been developed {\it ad hoc} to handle and analyze trajectories obtained from molecular simulations, and provide simple and efficient interfaces for facilitating the extraction of relevant structural and dynamical information out of them by integrating (more or less) standard descriptors. 

Together with widely used descriptors, such as number of neighbors, distances, etc., recently, new types of advanced descriptors have also been developed, which can provide information-rich characterizations of the systems under study. 
These can essentially be subdivided into two categories: static (time-independent) and dynamical (time-dependent) descriptors. 
For the former family, some relevant examples include, e.g., the Smooth Overlap of Atomic Positions (SOAP)\supercite{bartok2013representing}  and the Atomic Cluster Expansions (ACE)\supercite{Drautz2019ACE}, to name a few. Such advanced descriptors provide a high-dimensional embedding of the local structure, symmetry, and degree of order in the positions of neighbors around a central unit. The abstractness and generality of such descriptors offer the advantage of capturing many local structural details in the micro-domains that compose a system without heavy assumptions based, e.g., on prior knowledge of the system.

At the same time, pattern recognition approaches to detect dominant motifs in, e.g., SOAP datasets suffer in detecting rare events that have a negligible statistical weight in the dataset -- these are only sparsely observed, but that are often key for the physics of the system.
Particularly useful for detecting such important fluctuations are descriptors such as {\it Time}SOAP\supercite{caruso2023timesoap}, which tracks fluctuations of the SOAP spectra, capturing local structural changes, and the Local Environments and Neighbors Shuffling (LENS)\supercite{crippa2023detecting}, which allows the detection of local dynamical fluctuations (e.g., changes in fluidity or viscosity, permutations, neighbors exchange events, etc.).
Coupled with well-suited clustering approaches that make it possible to systematically discriminate statistically relevant fluctuations from noise and to classify them based on their similarity and diversity (e.g., Onion clustering\supercite{becchi2024layer, martino2025data}), these descriptors allow one to obtain deep insights into the local and collective dynamical events that may appear in a system\supercite{doria2025data}, and to study their correlations in space and time, thereby elucidating the mechanisms behind physical phenomena in an exquisitely data-driven way\supercite{caruso2025classification}.  

Thanks to their general and abstract character, such methods and approaches can, in principle, be applied to any kind of dynamically complex system. 
At the same time, all such advanced approaches use their own platform, codes or software, which has, until now, made their cross-application for the analysis of trajectories (whether obtained from simulations or experiments) substantially fragmented. To streamline the use of these methods and extend them beyond molecular simulations, we have developed and here describe {\tt dynsight}.

\begin{figure}[t]
    \centering
    \includegraphics[width=\linewidth]{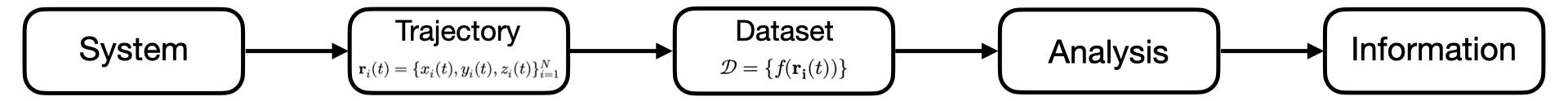}
    \caption{\textbf{The {\tt \textbf{dynsight}} workflow} Typical {\tt dynsight} pipeline, where trajectories, obtained from either simulations or experimental videos, are converted into datasets through the computation of descriptors, and subsequently analyzed to extract physically-relevant information.}
    \label{fig:pipeline}
\end{figure}

The typical workflow of a {\tt dynsight} analysis is schematically illustrated in Figure~\ref{fig:pipeline}. In particular, it starts with the extraction of the trajectories of the system under study. While this first step is straightforward in typical computer simulations (where the individual trajectories of all units moving in the system are automatically recorded at every simulation time step), obtaining microscopic-level trajectories from experimental data often remains challenging. 
To this end, we have implemented algorithms based on computer vision, object detection, and particle tracking that automatically and efficiently detect and track the individual units that move over time in experimental videos, providing as output their trajectories. Once the trajectories are available, {\tt dynsight} provides methods to compute a variety of advanced single-particle descriptors\supercite{crippa2023detecting, caruso2023timesoap, martino2025data}, and a flexible platform to eventually implement other desired ones.  
Useful tools to smooth signals and reduce noise in a physically meaningful way are also included\supercite{donkor2024beyond}, as well as an efficient unsupervised clustering algorithm for single-point time-series analyses that can detect the different fluctuations and microscopic dynamical domains that compose a system or signal (i.e., Onion Clustering\supercite{becchi2024layer}, although many other clustering algorithms can be easily interfaced or integrated within {\tt dynsight}). 

By streamlining extraction, preprocessing, clustering, and analysis in a flexible and broadly applicable way, {\tt dynsight} facilitates the understanding of datasets or signals, offering a robust and useful platform for advancing the study of complex systems and complex physical phenomena, as well as supporting the analysis of complex data and signals in general. The implementation of these methods in Python ensures seamless interaction and expandability with existing software, while the open-source nature of {\tt dynsight} enables straightforward integration of additional descriptors and analysis methods. 

\section{Software overview}

\subsection{How {\tt dynsight} unifies trajectory analysis}

In its current version (v2025.8.27), {\tt dynsight} is structured to support a wide range of tasks commonly encountered in the analysis of many-body dynamical systems. These tasks include handling trajectory data, computing single-particle descriptors, performing time-series clustering, and conducting various auxiliary analyses. To achieve this, {\tt dynsight} is organized into specialized modules, each addressing a specific aspect of this workflow (documentation describing their use is available at \href{https://dynsight.readthedocs.io/en/stable/trajectory.html}{dynsight.readthedocs.io/en/stable/trajectory}). 
Specifically, the {\tt dynsight.trajectory} module includes classes that streamline the entire analysis process into simple Python objects:

\paragraph{{\tt Class \textbf{Trj}}} This object contains a many-body trajectory, in the form of an {\tt MDAnalysis.Universe}\supercite{michaud2011mdanalysis}, and offers, among others, methods for the calculation of several single-particle descriptors (see Table~\ref{tab:tab1}). 

\paragraph{{\tt Class \textbf{Insight}}} This object contains a single-particle descriptor, in the form of a {\tt np.ndarray}\supercite{harris2020array} with shape {\tt (n\_atoms, n\_frames, n\_components)}. Methods include the computation of other descriptors, dimensionality-reduction techniques, and clustering algorithms (see Table~\ref{tab:tab1}), together with fast load from/save to file methods. 

\paragraph{{\tt Class \textbf{ClusterInsight}}} This object contains a clustering analysis on a single-particle descriptor, in the form of a {\tt np.ndarray}\supercite{harris2020array} of labels, with shape {\tt (n\_atoms, n\_frames)}. Methods include functions for plotting the clustering results, together with fast load from/save to file methods. 

\begin{table}[htbp]
 \begin{center}
 \caption{Selected list of {\tt dynsight.trajectory} methods. }
 \label{tab:tab1}
  \begin{tabular}{||l||l||}
   \hline
   \multirow{1}{*}{Method's type} & {Available methods} \\ \cline{2-2} \hline\hline
   \multirow{6}{*}{{\tt Trj} $\rightarrow$ {\tt Insight}}
    & {\tt Trj.get\_coord\_number()} \\ \cline{2-2} 
    & {\tt Trj.get\_coordinates()} \\ \cline{2-2} 
    & {\tt Trj.get\_lens()} \\ \cline{2-2} 
    & {\tt Trj.get\_soap()} \\ \cline{2-2} 
    & {\tt Trj.get\_timesoap()} \\ \cline{2-2} 
    & {\tt Trj.get\_orientational\_op()} \\ \cline{2-2} 
    & {\tt Trj.get\_velocity\_alignment()} \\ \hline
   \multirow{3}{*}{{\tt Insight} $\rightarrow$ {\tt Insight}}
    & {\tt Insight.get\_angular\_velocity()} \\ \cline{2-2} 
    & {\tt Insight.spatial\_average()} \\ \cline{2-2} 
    & {\tt Insight.get\_tica()} \\ \hline
   \multirow{2}{*}{{\tt Insight} $\rightarrow$ {\tt ClusterInsight}}
    & {\tt Insight.get\_onion()} \\ \cline{2-2}
    & {\tt Insight.get\_onion\_smooth()} \\ \hline
  \end{tabular}
 \end{center}
\end{table}

\subsection{Trajectory extraction from experimental videos: the {\tt vision} and {\tt track} modules}

\begin{figure}[H]
    \centering
    \includegraphics[width=\linewidth]{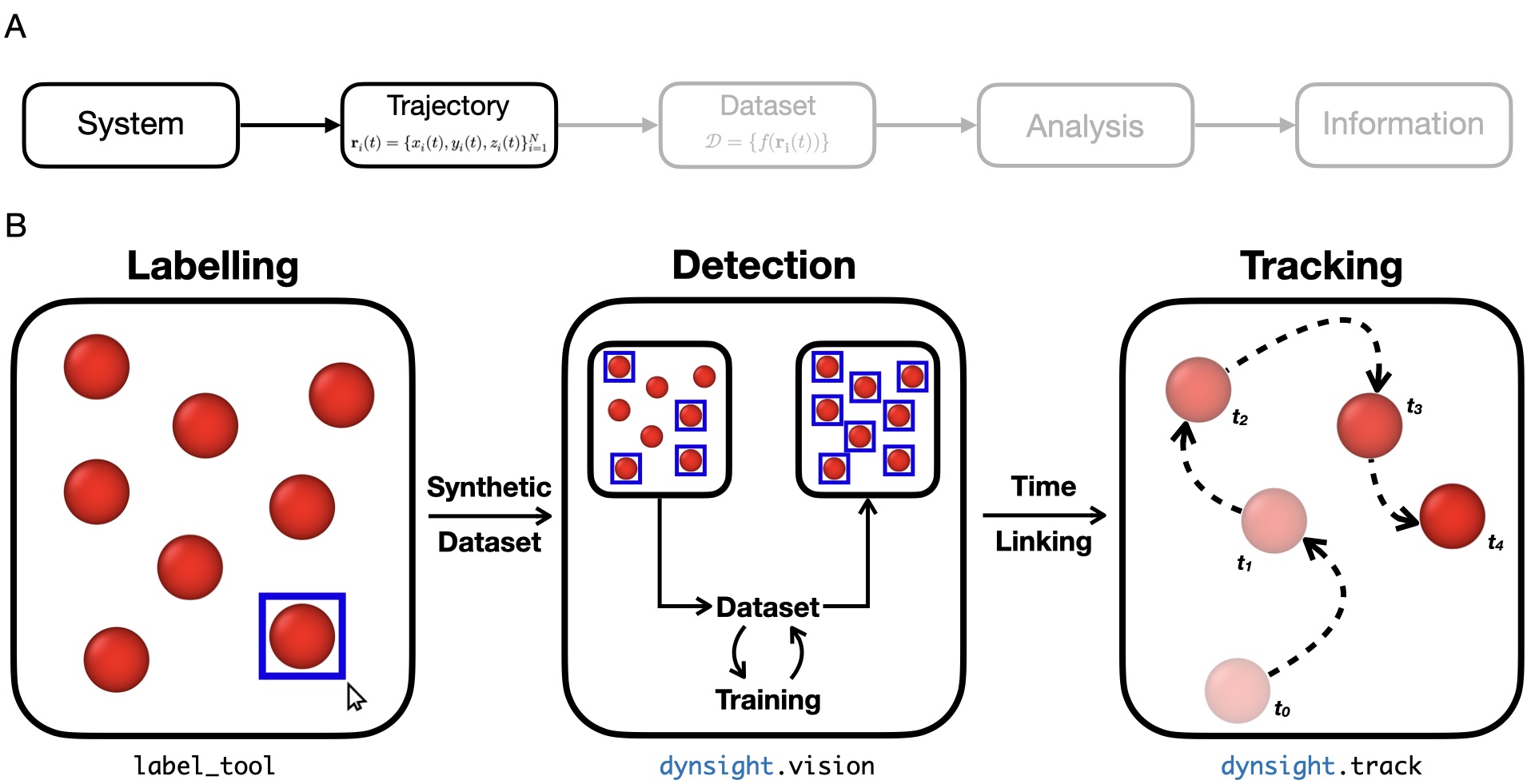}
    \caption{\textbf{The {\tt vision} and {\tt track} modules of {\tt dynsight}}. (A) A first operative phase aims to extract the trajectory data from experimental videos or simulations. (B) Schematic representation of the workflow to extract 2D trajectories from experimental videos. Procedure of labeling (left), training (center), and tracking (right) to obtain the trajectories of the moving objects/particles from input video in {\tt dynsight}.}
    \label{fig:vision}
\end{figure}

To uncover details about the motion and interactions of individual particles in a trajectory, we recognized the need to extract and track individual particles within trajectories, which may then be analyzed using descriptors.
While this is trivial in computer simulations, finding and tracking particles with high accuracy in experimental trajectories is an open challenge\supercite{zhang2015tracking, yao2020machine}.
To this end, we have implemented algorithms based on computer vision ({\tt dynsight.vision}) and tracking ({\tt dynsight.track}) that can automatically extract trajectories from experimental videos (see Figure~\ref{fig:vision}B). The {\tt dynsight.vision} module leverages convolutional neural network models from the widely used YOLO library\supercite{yolo11_ultralytics} to detect objects in experimental trajectory videos (i.e., from a set of consecutive snapshots or frames) and save the position of each detected item/object.
The {\tt vision} module provides a standalone web application called {\tt label\_tool} with a user-friendly graphical user interface (GUI) that allows the user to manually label an initial sample of the object that the user wants to detect (Figure~\ref{fig:vision}B, left). This preliminary annotation is used to automatically generate a synthetic dataset that facilitates the first training phase of the detection model. Subsequently, an iterative process of detection and retraining using the experimental video itself further refines the model to recognize the specific objects (Figure~\ref{fig:vision}B, center).

The {\tt dynsight.track} module applies the tracking algorithm {\tt trackpy}\supercite{trackpy} to assign a unique ID to each particle detected by the {\tt vision} module, enabling the reconstruction of individual trajectories across frames (Figure~\ref{fig:vision}B, right). In this current version of {\tt dynsight} (v2025.8.27), the {\tt dynsight.vision and \tt dynsight.track} modules have implemented these methods for single particle/object detection and tracking, nonetheless, given the flexibility and open character of the {\tt dynsight} platform, other detection and tracking algorithms/methods shall be eventually implemented/patched in the future.

\subsection{Computing single-particle descriptors}

{\tt dynsight} facilitates computation of several single-particle descriptors, which are functions of the particle trajectories ($x, y, z$ coordinates, velocities, etc.).
The descriptors already available in the current release (v2025.8.27) of {\tt dynsight} are listed below.

\begin{figure}[htbp]
 \centering
 \includegraphics[width=\linewidth]{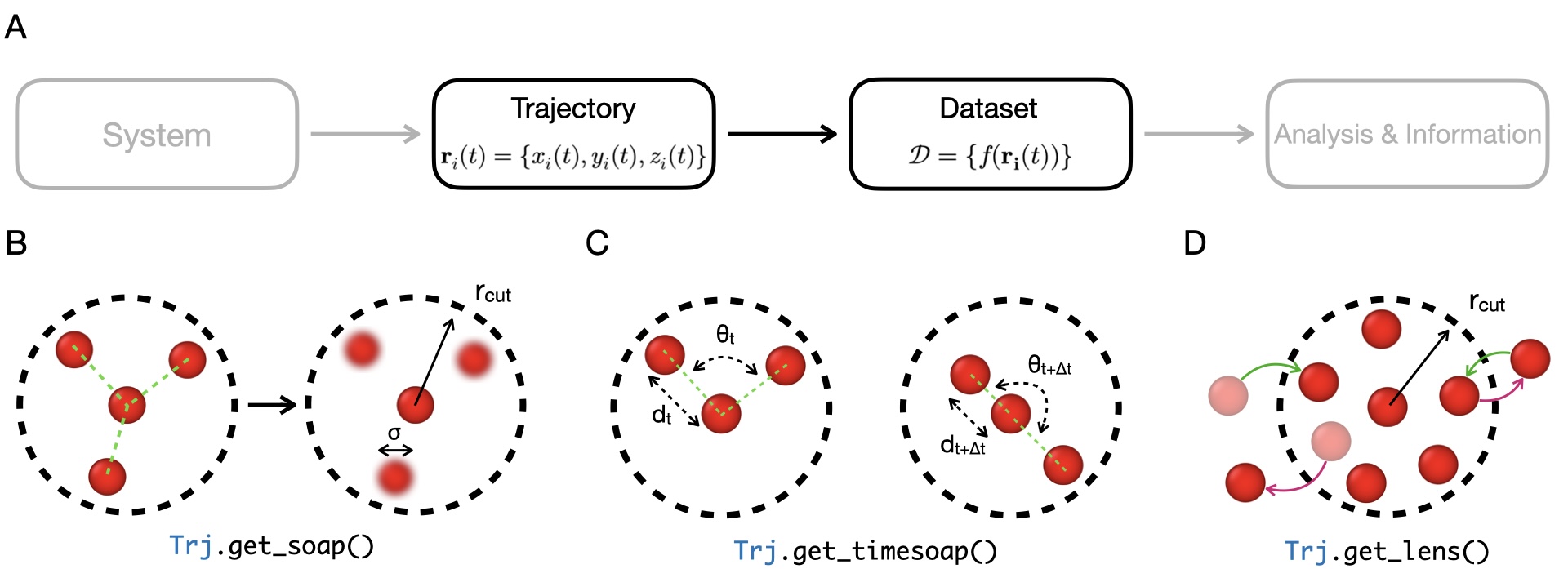}
 \caption{\textbf{Descriptors}: Example descriptors in {\tt dynsight} for analyzing trajectories to produce a dataset (A).
 (B) Smooth Overlap of Atomic Position (SOAP)\supercite{bartok2013representing}, a continuous, high-dimensional representation of the local particles' density around a particle. 
 (C) {\it time}SOAP\supercite{caruso2023timesoap}, a uni-dimensional measure of the temporal variation of the SOAP spectra.
 (D) Local Environments and Neighbors Shuffling LENS\supercite{crippa2023detecting}, a measure of the degree of neighbors' reshuffling around a particle.}
 \label{fig:descriptors}
\end{figure}

\paragraph{LENS: Local Environments and Neighbors Shuffling.}
LENS\supercite{crippa2023detecting} (see Figure~\ref{fig:descriptors}D) is a single-particle descriptor designed to analyze local dynamics in complex systems by tracking changes in the immediate surroundings of individual units over time, thereby capturing the degree of local dynamical rearrangement of the neighborhood of each unit. For details of the LENS descriptor and its applicability, we refer interested readers to the original paper\supercite{crippa2023detecting}. 
In short, LENS is a permutationally-variant, structurally-invariant descriptor: it is particularly well suited to capture local dynamical fluctuations (more so than, e.g., local structural transitions, for which other descriptors are better suited; see below) even in cases where these are local, collective, or rare events\supercite{crippa2023detecting}.
LENS has proven particularly useful to identify and classify local fluctuations and dynamic domains, such as distinguishing between solid-like and liquid-like regions in complex systems or identifying interfacial dynamics in phase transitions\supercite{crippa2023detecting}, and in elucidating key mechanisms behind non-trivial dynamics\supercite{caruso2025classification}. With {\tt dynsight}, LENS can be easily computed from a many-body trajectory using the {\tt Trj.get\_lens()} method (see Figure~\ref{fig:descriptors}D).

\paragraph{SOAP: Smooth Overlap of Atomic Positions.}
The SOAP\supercite{bartok2013representing} (see Figure~\ref{fig:descriptors}B) descriptor, is a structural descriptor that provides a high-dimensional representation of the local structure around a particle by encoding the relative spatial arrangement of neighboring particles into a smooth and continuous representation. In this sense, the SOAP power spectrum serves as a descriptor of the degree of local order or disorder in the relative displacements of the weights around a center (symmetry, distances, etc.). For details concerning the SOAP descriptor, we refer interested readers to the original paper\supercite{bartok2013representing} or to dedicated works focusing on the analysis of SOAP dataset\supercite{Gasparotto2020Identifying, Capelli2022Ephemeral, deringer2018computational, monserrat2020liquid} or time-series data\supercite{Lionello2025Relevant, martino2025data}. 
{\tt dynsight} uses the {\tt DScribe} package\supercite{dscribe} to perform SOAP calculations, making it easier to streamline these calculations within a single Python workflow, via the {\tt Trj.get\_soap()} method. 

\paragraph{{\it Time}SOAP: Tracking time-variations in SOAP spectra.}
{\it Time}SOAP\supercite{caruso2023timesoap} ($t$SOAP, see Figure~\ref{fig:descriptors}C) is a time-dependent descriptor that, starting from the structural description of local environments provided by SOAP, detects and tracks high-dimensional fluctuations over time in the SOAP spectra. It captures local structural changes or transitions in the neighborhood of every unit.
In this way, $t$SOAP enables the identification of critical events or a microscopic-level characterization of complex phenomena such as phase coexistence and structural transformations\supercite{caruso2023timesoap} that may be lost in standard pattern recognition analyses of SOAP datasets or when using simpler metrics. 
$t$SOAP is a structurally-variant, permutationally-invariant descriptor: in this sense, it is complementary to LENS, and well suited to capture local structural fluctuations \supercite{caruso2025classification}. For more details on $t$SOAP descriptor, we refer interested readers to the dedicated papers\supercite{caruso2023timesoap, caruso2025classification}. 

With {\tt dynsight}, {\it Time}SOAP can be computed from a SOAP spectrum time-series using the {\tt Insight.get\_angular\_velocity()} method or the {\tt dynsight.soap.timesoap()} function (see Figure~\ref{fig:descriptors}C). 

\paragraph{Miscellaneous descriptors.}
The {\tt dynsight.Trj} classes offer methods to compute, from a many-body trajectory, the number of neighbor particles, velocity, local velocity alignment\supercite{vicsek2012collective}, and the orientational order parameter\supercite{aeppli1984hexatic}.
Moreover, the Time-lagged Independent Component Analysis (tICA)\supercite{molgedey1994separation} dimensionality reduction algorithm is also implemented using {\tt deeptime}\supercite{hoffmann2021deeptime}, to simplify the treatment of high-dimensional descriptors. 
It is worth noting that, while such descriptors were originally developed for atomistic or molecular systems, their abstract character makes them well suited to study systems at virtually any scale.

In the {\tt dynsight} framework, adding new descriptors or {\tt Insight}s is simplified for users.
The computation of any scalar or vectorial single-particle quantity can be easily implemented within {\tt dynsight}, making it a viable platform for method development. 

\subsection{Unsupervised clustering for single-point time-series analysis: e.g., {\it Onion Clustering}}
The current version of {\tt dynsight} (v2025.8.27) implements {\it Onion Clustering}\supercite{becchi2024layer} as an efficient unsupervised clustering method for single-point time-series analysis. Notably, the flexibility and open character of the {\tt dynsight} platform make it easy to also implement or interface other clustering methods that can then be used to analyze the trajectories and datasets obtained when using the other modules.

\begin{figure}[h]
    \centering
    \includegraphics[width=\linewidth]{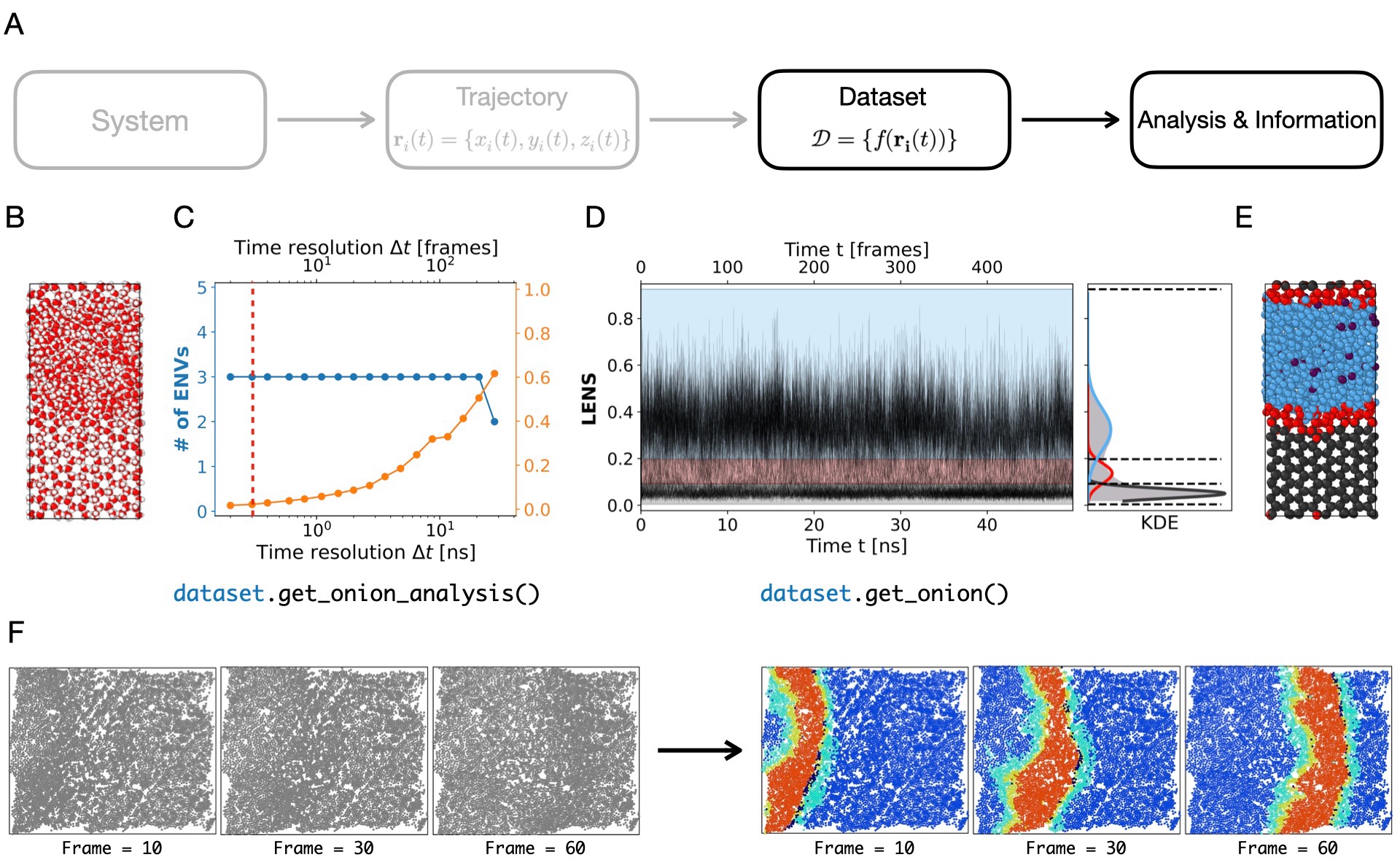}
    \caption{\textbf{Time-series clustering}: Example of {\tt dynsight} usage for information extraction through the {\it Onion Clustering} algorithm from descriptor based datasets (A). (B) Prototypical example of a water-ice molecular dynamics simulation. (C) Number of identified environments (blue) and unclassified data (orange) as a function of $\Delta t$ used by {\it Onion Clustering}. (D) Time-series of the LENS descriptor classified by {\it Onion Clustering}, with corresponding kernel density estimation (KDE) on the right. (E) Snapshot colored according to the detected clusters. Application of {\it Onion Clustering} to an experimental video.}
    \label{fig:onion}
\end{figure}

For a detailed description of the {\it Onion Clustering} method, we refer interested readers to the original paper\supercite{becchi2024layer} and to other related applied works that clearly explain the potential and advantages of this method\supercite{doria2025data, becchi2025maximum, martino2025data, Lionello2025Relevant}. 
In short, {\it Onion Clustering} is an efficient algorithm for single-point clustering of time-series data that allows identifying and classifying all fluctuations and local dynamical domains (including sparse or hidden ones, which may be obscured by the noise of the dominant ones) that can be characterized in a statistically robust manner as a function of the temporal resolution $\Delta t$ used in the analysis. 
More in detail, {\it Onion Clustering} performs a series of clustering analyses, each with a different time resolution $\Delta t$ (i.e., different minimum lifetime required for the identified states), and outputs the microscopic dynamical domains that can be effectively distinguished as a function of the $\Delta t$ (the typical output is the so-called {\it Onion plot}: Figure~\ref{fig:onion}C). 
The clustering exploits an iterative procedure where, at the end of the process, each data point is either classified in one of the identified states, or labeled as ``unclassified'' at the selected time resolution (Figure~\ref{fig:onion}C: orange -- the fluctuations occurring faster than the time resolution used). 
Performing this analysis at different values of the time resolution $\Delta t$ allows automatic identification of the optimal choice of $\Delta t$, the one that maximizing the statistically robust subdivision of the time-series data in the largest number of microscopic environments (Figure~\ref{fig:onion}C,D): i.e., this is the time resolution that guarantees maximum information extraction from the data\supercite{becchi2025maximum, doria2025data}. 

Onion clustering can be performed on virtually any time-series using the {\tt Insight.get\_onion()} method and its variations. Note that while this method offers considerable advantages in statistical robustness and interpretability\supercite{becchi2024layer, becchi2025maximum}, many other clustering methods can be implemented or interfaced with other modules of {\tt dynsight}, thanks to the open character of this platform.

\subsection{Miscellaneous modules useful for data analysis: the {\tt analysis} module}
The {\tt dynsight.analysis} module provides other key tools for pre- and post-processing and statistical analysis of many-body trajectories, including: 
\begin{itemize}
    \item Denoising tools: the present version (v2025.8.27) of {\tt dynsight} implements a spatial denoising approach\supercite{donkor2024beyond}, which can be applied to time-series computed from trajectories in order to enhance their signal-to-noise ratio. Noise filtering tools are currently being optimized in our group, which will also be made available in {\tt dynsight}.
    \item The computation of the radial distribution function $g(r)$ from many-body trajectories, which quantifies how particle density varies as a function of distance from a reference particle. 
    \item The computation of several information-theory quantities such as e.g., Shannon entropy\supercite{shannon1948mathematical}, Sample entropy\supercite{richman2000physiological, richman2004sample}, and information gain achieved by clustering algorithms\supercite{becchi2025maximum}, which allow a quantitative evaluation of the information content and extraction during the analyses. 
\end{itemize}

While these additional tools are included in the present version of {\tt dynsight} (v2025.8.27), the open character of this platform makes it straightforward to implement, interface with, or patch other utilities that may be needed.

\subsection{Fostering and facilitating open-data archiving}
By design, {\tt dynsight} is a platform that promotes transparent data management and open data practices. We have implemented an integrated logging module that records a human-readable log of all performed analyses. The {\tt dynsight.logger} can also automatically generate an archive containing all the relevant data and metadata. This standardized archive is thus immediately ready for deposition in common open data repositories (e.g., {\it Zenodo}), which comply with the FAIR principles (Findable, Accessible, Interoperable, Reusable).
This feature is useful not only for tracking the analysis workflow but also for facilitating the reproducibility of results, ultimately contributing to the advancement of research within the community.

\section{How to use {\tt dynsight}}
{\tt dynsight} is available on the PyPI repository at~\href{https://pypi.org/project/dynsight/}{pypi.org/project/dynsight} and can be installed with {\tt pip install dynsight}. The code is open-source under the MIT license and the documentation is available at \href{https://dynsight.readthedocs.io}{dynsight.readthedocs.io/en/latest}.

\begin{figure}[H]
\centering
\includegraphics[width=\linewidth]{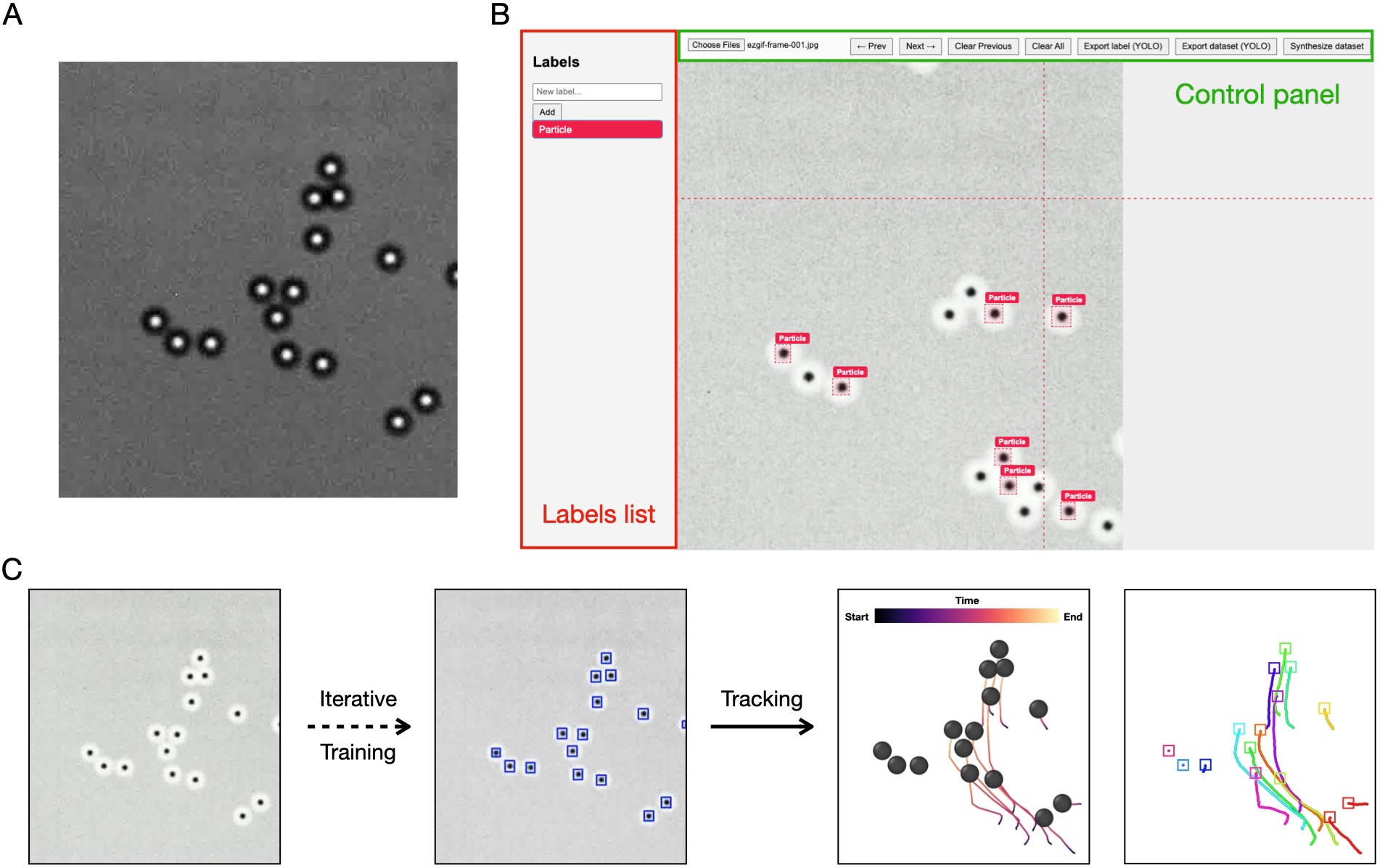}
\caption{\textbf{Workflow of {\tt Label\_tool} and {\tt vision} / {\tt track} modules.} (A) Single frame taken from a representative experimental video of an example complex colloidal system\supercite{Liu2021Activity}. For visual purposes, a relatively simple many-body system is used here as a representative example, while it is worth noting that such approaches become particularly useful when the complexity of the system increases\supercite{becchi2024layer, caruso2025classification, doria2025data, Lionello2025Relevant}. (B) The {\tt Label\_tool} GUI allows the user to manually select prototypical examples of target particles from a color-corrected frame. (C) Using these annotations, {\tt dynsight.vision} performs iterative training and detection to locate all particles across frames, while {\tt dynsight.track} associates each detection with its corresponding trajectory.}
\label{fig:label_tool}
\end{figure}

\paragraph{The {\tt Label\_tool} GUI of the {\tt vision} module.}
Figure~\ref{fig:label_tool} shows the integration of the {\tt label\_tool} application with {\tt dynsight.vision} and {\tt dynsight.track} for particle trajectory extraction from experimental videos. Here, we used a preprocessed Quincke rollers experimental video\supercite{Liu2021Activity} (see Figure~\ref{fig:label_tool} A), where color correction was applied to enhance particle visibility while suppressing background noise. The {\tt label\_tool} graphical user interface (GUI) enables users to efficiently create initial training data by selecting representative examples of target particles directly from single video frames. These user-provided annotations are then automatically expanded into a synthetic dataset by randomly redistributing the labeled particles over multiple artificial images, effectively increasing the diversity and robustness of the training dataset. As shown in Figure~\ref{fig:vision}B, this synthetic dataset serves as the initial input for {\tt dynsight.vision}, which performs iterative learning and detection to identify all bounding boxes corresponding to particles across the entire frame sequence. The detected positions are subsequently processed by {\tt dynsight.track}, which associates detections over consecutive frames to reconstruct the full trajectories of individual particles (see Figure~\ref{fig:label_tool}C). This workflow enables the semi-automated generation of accurate and reproducible tracking data, providing a robust starting point for the analysis methods presented above.

\paragraph{An example of a typical {\tt dynsight} analysis workflow.}
The first step is usually to create a {\tt trajectory.Trj} object from some trajectory files. In this example, we are using a water/ice coexistence trajectory stored in the {\tt dynsight} example folder (see Figure~\ref{fig:onion}B).
\begin{lstlisting}[language=Python, caption=Loading trajectory files in {\tt dynsight}.]
from pathlib import Path
from dynsight.trajectory import Trj

files_path = Path("dynsight/examples/analysis_workflow")
trj = Trj.init_from_xtc(
    traj_file=files_path / "oxygens.xtc",
    topo_file=files_path / "oxygens.gro",
)
\end{lstlisting}
The variable {\tt trj} now contains the trajectory (as an {\tt MDAnalysis.Universe}, also allowing analysis using that software), and using the methods of the {\tt trajectory.Trj} class we can perform all the desired analyses on this trajectory. For instance, we can compute compute LENS (see Figure~\ref{fig:onion}D):

\begin{lstlisting}[language=Python,
  caption={Compute the LENS descriptor with a cutoff radius $r_\text{cut}=7.5$~\AA.}]
  neighbors_count, lens = trj.get_lens(r_cut=7.5)
\end{lstlisting}

Note that the units used for the calculation of the various descriptors depend on the units used in the input trajectory. We refer readers interested in data-driven methods to assess optimal space and time analysis resolutions to dedicated papers on this topic\supercite{doria2025data, becchi2025maximum}. 

The method {\tt Trj.get\_lens()} returns a {\tt trajectory.Insight} object, which in its {\tt .dataset} attribute contains the LENS values computed on the {\tt trj} trajectory. The {\tt trajectory.Insight} class offers its own methods for further analysis. For instance, one can perform spatial averaging\supercite{donkor2024beyond, martino2025data} of the LENS values: 

\begin{lstlisting}[language=Python, caption=Compute spatial average as an example of possible data processing.]
trj_lens = trj.with_slice(slice(0, -1, 1))
lens_smooth = lens.spatial_average(
    trj=trj_lens,
    r_cut=7.5,
    num_processes=6,
)
\end{lstlisting}
Note that, since LENS is computed for intervals between frames, we need to use a sliced trajectory, which we get with the {\tt Trj.with\_slice()} method. 
Finally, we can perform clustering on the {\tt lens\_smooth.dataset}, using for instance the {\tt Insight.get\_onion\_smooth()} method (see Figure~\ref{fig:onion}C-F), and plot the results:

\begin{lstlisting}[language=Python, caption=Onion Clustering computing and plotting]
lens_onion = lens_smooth.get_onion_smooth(delta_t=10)

lens_onion.plot_output(
    file_path=files_path / "tmp_fig1.png",
    data_insight=lens_smooth,
)

lens_onion.dump_colored_trj(
    trj=trj_lens,
    file_path=files_path / "colored_trj.xyz",
)
\end{lstlisting}
\paragraph{Examples and recipes.}
Example scripts included in this implementation of {\tt dynsight} are designed to help users smoothly integrate the {\tt dynsight} tools into their own code. These are provided as open-access in the documentation and include, e.g., trajectory extraction from videos, computation of single-particle descriptors, use of onion clustering, and selection of the optimal spatiotemporal analysis resolution\supercite{doria2025data, becchi2025maximum}.

\paragraph{How to contribute.}
The authors welcome any contribution to improve and expand the {\tt dynsight} code and its modules.
Any user can submit comments, requests, or additions by opening an issue or submitting a pull request in the {\tt dynsight} GitHub repository~\href{https://github.com/GMPavanLab/dynsight}{github.com/GMPavanLab/dynsight}. 

\section{Conclusions}
We have presented {\tt dynsight}, an open Python platform for resolving, processing, and analyzing trajectory data from both experimental and simulated systems.
By combining trajectory extraction, descriptor computation, and unsupervised clustering methods in a single platform, {\tt dynsight} streamlines workflows that would otherwise require substantial expertise and familiarity with diverse environments/tools.
The flexible design of the platform allows application across diverse systems and scales, enabling detailed microscopic-level analyses as well as broader studies of local and collective dynamics and expected behaviors.
Overall, {\tt dynsight} lowers the barrier to trajectory-based analysis, promoting reproducibility, consistency, and wider adoption of advanced dynamic descriptors.
The possibility of using a single common platform, as well as the same methods and analysis tools, to study systems from atomic simulations to macroscopic experimental setups also opens unique opportunities to study complex behaviors and emergent properties through different scales in virtually any complex system for which a trajectory of the constitutive units can be resolved. We believe that this will constitute a flexible and solid foundation for advancing the study of complex systems as well as the analysis of static and dynamic data (or signals) in general.

\section*{Acknowledgments}
The authors acknowledge the funding received by the European Research Council under the European Union’s Horizon 2020 research and innovation program (Grant Agreement No. 818776-DYNAPOL, to G.M.P.). A.T. also acknowledges the support received by the European Union under the Next Generation EU program (Mission 4 Component 1 CUP E13C22002930006).
The authors acknowledge (in alphabetic order) Lucrezia Baldo, Cristina Caruso, Matteo Cioni, Martina Crippa, Massimo Delle Piane, Domiziano Doria and Chiara Lionello for the useful testing, review, and feedback.

\printbibliography

\end{document}